\documentclass{article}
\usepackage{amssymb}

\newtheorem{theorem}{Theorem}[section]
\newtheorem{lemma}{Lemma}[section]
\newtheorem{corollary}{Corollary}[section]

\title{A Zeno Story\footnote{Written evening, 11-26-01, in Delphi,
Greece, during the 22nd Solvay Conference on Physics, following a
lecture that day on Quantum Zeno Effect by Professor S. Pascazio.  A 
substantial portion of this conference dealt directly or indirectly with
the Zeno effect, and I was struck by the fact that some important
mathematical beginnings of the Zeno story were not known to many
participants.  Professor Pascazio, however, was clearly cognizant of some
of the early history.  Therefore I wrote this draft (in preliminary form)
and the next day circulated it to Professor Pascazio and a few others at
the conference.  Also in that draft I presented my Zeno alternative,
worked out a few days prior (on 11-18-01, to be precise) to the Solvay
Conference.  Professor Pascazio urged me to publish this account.
Professor Antoniou has long urged me to publish some account of my early
work on the Zeno problem.}}
\author{Karl Gustafson\\
Department of Mathematics, University of Colorado,\\
Boulder, CO 80309--0395 USA}

\begin{document}
\maketitle

\begin{abstract}
I describe the early (1974--75) work I did on what is now called the Zeno
problem in quantum mechanics.  Then I propose a new formulation which
may obviate a vexing problem of operator limits and which also may be
more measurement-compatible.
\end{abstract}

\section{Introduction}
I was involved in the formulation of the quantum Zeno effect some years
ago, in 1974--75, working with B. Misra in Boulder.  I wrote up some of
that in [1] at that time, as a first draft of a paper.  I also lectured
on the problem at the Rocky Mt. Mathematics Consortium summer school in
Bozeman in summer 1975 [2].  These contributions were never published,
although I did summarize briefly some of the issues in [3].  Therefore I
would like to return to this problem, describe the situation in 1974--75,
and also propose a new formulation of the Zeno problem, using my recent
results in [4].

The Zeno's paradox ``a watched pot never boils'' has been presented in
[5] and has since been discussed by many authors.  As I recall, the
problem was presented to Misra and me by Professor Josef Jauch when he
visited us in Boulder in April 1974.  In my work on the problem, I
preferred to call it ``The Counter Problem'' [1].  I had nothing to do
with the later and more popular formulation as ``Zeno's Paradox'', although
I note now that in [2] I did state 
\newpage

``When 
$\limsup \Vert (PU_{t/n} P)^n \psi\Vert \stackrel{a.e.}{=} \Vert
P\psi\Vert^2$,
then $\mathbb P$ (affirmative answer) $= {\mathbb P}$ (that part was in
$C$ initially) $\sim$ ``particle cannot decay''.  Here $C$ refers to a counter
which can be visualized as absorbing the wave packet.  In fact we were mostly
following the formulation of Friedman [6], who deserves more citation.
Like Friedman, I could not resolve the main mathematical issue, which was
the existence and properties of an operator limit
$$s\mbox{-}{\lim_{n\rightarrow \infty}} (PU(t/n)P)^n \eqno(1.1)$$
In (1.1) $U$ is a unitary evolution group and $P$ is an orthogonal
projection on a Hilbert space $\mathcal H$.  As [1] and [2] reveal, I
spent considerable time on this question, because it seems to represent
an important new domain in the theory of unbounded operator limits.
Briefly, neither the ``Trotter'' theory of operator limits found in [7],
nor the monotone forms  approach found in [8], could be made to apply to
the limit (1.1).  Misra and I did publish the two related papers [9,10]
but (frankly) I was surprised when [5] appeared, because I thought the
problem was not solved.

As it turns out, the publication of [5] created considerable interest, so
that was a good thing for physics.  However [5] is based on never-proven
mathematical assumptions.  ``Paradoxes'' are often incompletely-cast
models, although how to most-correctly cast a situation may not be easy.

\section{The Continual Observation Problem}
Friedman [6] treated what is now called the Zeno problem by reducing it
to the question (1.1) above and to related operator-theoretic
considerations.  I quote ([6], Section 5, p. 1007)

\begin{quotation}
My original motivation for studying semigroup product formulas was the
following situation:  $P^t$ is a contraction semigroup on the Hilbert
space $\mathcal H$, $E$ an orthogonal projection in $\mathcal H$.  If
strong $\lim_{n\rightarrow\infty} (EP^{t/n} E)^n$ exists, we call it the
compression of $P^t$ by $E$.  Formally one expects the limit to be $\exp
(tEAE)$ where $A$ is the infinitesimal generator of $P^t$.  However, $EAE$
is not in general the generator of a contraction semigroup, nor is its
closure.
\end{quotation}

\noindent
Friedman [6] goes on to give some partial results when the infinitesimal
generator $A$ is a negative selfadjoint operator, but he does not resolve
the strong limit question (1.1).  In ([6], Section 6) he then goes to the
quantum mechanical continual observation problem.  First he discusses
Feynman's ideal measurement formulation, that of determining whether or
not a trajectory of a particle lies in a given space-time region.  Then
he presents a modified formulation of ideal measurements attributed to Ed
Nelson, which becomes ([6], p. 1010) 

\begin{quotation}
This makes it reasonable to define
$$\lim_{n\rightarrow\infty} \Vert (E\exp (-itH_0 /n)E)^n \psi (x)\Vert^2
\eqno(2.1)$$
(if it exists) as the probability that---an ideal measurment of
`continual observation' during the interval $(0,t)$ for the purpose of
determining whether the particle stays in $\mathcal E$ yields the result
that the particle is indeed constantly in $\mathcal E$ during $(0,t)$.
\end{quotation}

I leave it to the interested reader to pursue further details in [6].  I
preferred to call the problem that of a (opaque) counter in [1,2,3].  The
Zeno interpretation was put in in [5].  Since then there is a large
literature.  For such, I just defer to [11,12,13] and citations therein.

Friedman [6] concludes:  

\begin{quotation}
The question of whether strong
$\lim_{n\rightarrow\infty} (E\exp (itH_0 /n)E)^n$ exists in $\mathcal
L^2$ and whether it is a unitary group remains open; to answer this would
seem to require a much deeper knowledge of semigroup product formulas.''
\end{quotation}

\noindent Very recently I have found [14] an alternative approach to this 
question which I now present in the next section.

\section{A new formulation}
Some recent authors [13] have rediscovered Friedman's work [6] or have
otherwise been led to wonder [11] if one cannot stay in the reversible
regime (e.g., lecture on 11-26-01 at the Solvay Conference by S.
Pascazio).  Although a bit oversimplified, this would mean that one could
say under general conditions that the operator limit (1.1) would be
$e^{iPHPt}$ where $H$ was the original self-adjoint infinitesimal
generator of the evolution group $U_t = e^{iHt}$.  Therefore one
naturally poses the question:  when is $PHP$ selfadjoint, i.e., when is
$(PHP)^* = PHP$?  Of course one asked this question in the beginning
[6,1,2,3].  The difficulty is that the usual operator-theoretic
conditions for product adjoints do not apply well to such projected
Hamiltonians.  Briefly, the usual conditions (e.g., see [15]) for
operators $A$ and $B$ to satisfy $(BA)^* = A^* B^*$ are either $A$ and
$B$ bounded, or if not, $A$ and $B$ Fredholm (closed ranges and finite
index).  The tacit assumption is, of course, that generally $A$ and $B$
are regarded as noncommuting operators.

What I propose [14] is to reformulate (1.1) as the question
$$(AHA)^* = A^* H^* A^* \eqno(3.1)$$
where $A$ replaces $P$.  What I have in mind is that the spatial
(Dirichlet) projection $P$ used by Friedman [6], see also [13], may not
be the best way to formulate a ``Zeno'' (continual) measurement.  Not
only will (3.1) be provable mathematically (below), but it also opens up
the use of new physical observables $A$ representing the act of
measurement.

Here is an example of my approach.  In [4] I recently obtained the 
following result.

\begin{lemma}
\mbox{[4]}
Let $A$ and $B$ be arbitrary densely defined operators in a Hilbert space
$\mathcal H$ and suppose the domain $\mathcal D (AB)$ is dense, the range
$\mathcal R (B) \supset \mathcal D (A)$, the domain $\mathcal D (B^*) 
\supset \mathcal R
(A^*)$, and that $B$ is 1--1.  Then $(AB)^* = B^* A^*$.  In particular,
when $A$ and $B$ are selfadjoint, then the conditions are $\mathcal D
(AB)$ dense, $\mathcal R (B) \supset \mathcal D (A)$, $\mathcal D (B)
\supset \mathcal R (A)$.
\end{lemma}

We may now apply this Lemma to the ``Zeno'' problem.  For simplicity, let
me first restrict here to observables $A$ which are bounded selfadjoint
operators on $\mathcal H$.  This stipulation may be relaxed later.
Then we have
$$(AHA)^* = (HA)^* A \supseteq (A^* H^*)A = AHA \eqno(3.2)$$
The relations in (3.2) follow because $A = A^* \in \mathcal B (\mathcal
H)$, I have assumed the domain $\mathcal D (HA)$ to be dense so that the
adjoint operator $(HA)^*$ exists, and one always has the relation $(AB)^*
\supseteq B^* A^*$ when all adjoint operators spoken of therein are
defined.  Thus the only remaining issue is to make the second relation in
(3.2) an equality.  To that end one may use the result of [4].  Therefore
I have shown the following.

\begin{theorem}
Let $A$ be a ``continual measurement observable'' such that $A = A^* \in 
\mathcal B
(\mathcal H)$, $\mathcal R (A) \supset \mathcal D (H)$, $\mathcal D (HA)$
dense.  Then $(AHA)^*$ is selfadjoint and the exponentiation $e^{iAHAt}$
is unitary.
\end{theorem}

Note that the condition $\mathcal R (A) \supset \mathcal D (H)$ in the
Theorem seems reasonable for ``continual'' measuring observables:  we
should have $A$ able to cover all possible wave functions $\psi$ in the
domain of the Hamiltonian.  Another way to view this requirement is that $A$
as measuror is also compatible with $A$ as preparor of state.  The condition 
$\mathcal D (HA)$ dense is a
technical one which could be loosely viewed as assuring that the
measurement observable $A$ is not ``too'' incompatible with the
Hamiltonian $H$ whose action is to be measured.  One could call these
measuring observables those of `full measurement'.

\section{Elaboration (after the conference)}
Here I would like to add a few clarifications to the above.
Further development of this approach will be pursued elsewhere [16].

\subsection{Measurement and Domains}
The issues surrounding measurement in quantum mechanics are numerous and
longstanding.  See for example [11, 17--20] for recent treatments.  Here I
will just make a few comments related to this paper.

Friedman [6], like many others, equates observation with measurement.  The
space--time region is $(0,t)\times$ Counter.  His main example, although
stated as a free particle evolution $e^{-itH_0} \psi_0$ where $H_0 =
-\sum_{i=1}^3 \partial^2 /\partial x_i^2$  is the three dimensional Laplacian
operator, is almost classical.  What I call the Counter is in [6] a 
closed bounded
three-dimensional real domain $\mathcal E$ with smooth boundary $\partial
\mathcal E$.  Variations of this example are studied in [12,13].

Now I would like to make an observation about this setting.  One of the
difficulties of Friedman's [6] model and analysis is that there are two
different operators posed, both called $H_0$.  The first $H_0$ is the free
space Laplacian $\Delta$ in $\mathcal L^2 (R^3)$.  Its domain $\mathcal D
(H_0)$, in order that it be selfadjoint, may be specified [21] by appropriate
Sobolev derivatives and appropriate decay rates at infinity.  Then $\mathcal
L^2 (R^3)$ is projected to $\mathcal L^2$ (Counter).  Friedman's projection
$E$ is implemented by multiplication by the characteristic function $\chi
(\mathcal E)$.  Within $\mathcal L^2 (\mathcal E)$ we then find another 
selfadjoint operator $H_0$, the Laplacian $\Delta$ in $\mathcal L^2 (\mathcal 
E)$ with specified Dirichlet boundary conditions.  The domain $\mathcal D
(H_0)$ of this operator is well known (see [21] and citations therein) but the
trace nature of what means $\psi = 0$ on the boundary $\partial \mathcal E$
for $\phi$ in $\mathcal D (H_0)$ is as is well known a delicate matter.
Remember that $\psi$ needed two Sobolev derivatives in the interior of
$\mathcal E$.  So to project the larger ``quantum mechanical'' evolution
$e^{-itH_0} \psi_0$ simply by multiplying by the characteristic function $\chi
(\mathcal E)$ does nothing at all about attending to the matter of the
regularity at the boundary needed by functions $\psi$ which are to be in the
domain $\mathcal D (H_0)$ of the other operator $H_0$, namely, the
Poisson--Dirichlet operator in $\mathcal L^2 (\mathcal E)$.  This, in my
opinion, lies at the root of Friedman's [6] inability to prove any rigorous
result about continual measurement.  Also, in my opinion, when Misra and
Sudarshan must assume [5, eqn (14)] $s\mbox{-}{\lim_{t\rightarrow O^+}} T(t) =
E$, one is on unfirm ground of the same underlying nature.  Stated another way
and more generally, we once again find the quantum-classical interface between
quantum probability wave and classical measuror, a delicate business.

On the other hand, the quantum mechanical evolutions $e^{-it\Delta}$
freespace and the classical evolution $e^{-it\Delta}$
Poisson--Dirichlet proceed easily on their way.  The former is analyzed in
[13] in one dimension and it is shown that, for initial state $\psi_0$ taken
with support in $\mathcal E = [0,1]$, the evolution corresponds to the
propagator of a particle in a square well with Dirichlet boundary conditions.
In [13] ``We prepare a particle in a state with support in $A$'' (we may take
$A \equiv \mathcal E \equiv [0,1] \equiv$ the counter) and this state $\psi_0$
is evolved by the free space evolution and frequently projected by $E =
\chi (\mathcal E)$
according to Friedman's prescription.  Much of the work in [13]
then consists in using the explicit, Green's function representation of the
evolution $e^{-itp^2 /2m}$ to show Zeno nondecay.  In this way they [13] are
able to obviate a use of Trotter formula approach [12] which we have seen many
times [1,2,6,7,12] does not quite go through.

Here I would like to make a second observation.  Consider any Dirichlet domain
$\mathcal E$ in $\mathcal L^2 (R^n)$.  By this I mean one for which the
Poisson--Dirichlet problem [21]
$$\Delta \psi = f \quad \mbox{in} \quad \mathcal E, \quad \psi = 0 \quad
\mbox{on} \quad \partial \mathcal E \eqno(4.1)$$
is well-posed.  Let $H_0$ denote the corresponding selfadjoint operator in
$\mathcal L^2 (\mathcal E)$. 
Consider any initial $\psi_0$ in $\mathcal D (H_0)$.  We notice that
means that $\psi_0 = 0$ on $\partial \mathcal E$ and also that $\psi_0$
possesses the necessary regularity, especially up to the boundary, to be in
$\mathcal D (H_0)$.  Now we know rather generally (e.g., [8, Chapter IX]) that
exponentiations $U_t$ commute with their infinitesimal generators, i.e., $U_t
H_0 \subseteq H_0 U_t$.  This means that $U_t$ maps $\mathcal D (H_0)$ into
$\mathcal D (H_0)$.  So the evolving Poisson--Dirichlet wave packet
$e^{-itH_0} \psi_0$ keeps vanishing on the surface of the counter.  Moreover
it is a unitary evolution in $\mathcal L^2 (\mathcal E)$.  This means that had
we started  with the free space evolution $e^{-it\Delta}$ and any initial
state $\psi_0$ in $\mathcal L^2 (R^n)$, then if we want to `count it' in the
way desired by Friedman [6] et al., we want a ``projection'' $E$ which maps us
not only to the counter $\mathcal E$ but also to $\mathcal D (H_0)$ where
$H_0$ is the Poisson--Dirichlet Hamiltonian.  That means restricting $\mathcal
L^2 (R^n)$ to $\mathcal L^2 (\mathcal E)  \cap \mathcal D (H_0)$.  In
other words, once you `project' into the counter, also satsifying the
Dirichlet trace boundary condition on the surface of the counter, then that
portion of your original wave-packet free particle evolution continues as a
unitary evolution within the counter.  Once you are in the counter, you stay
in the counter always.  This, in my opinion, is the real meaning of the claim
in [13] that ``Zeno dynamics uniquely determines the boundary conditions, and
they turn out to be of Dirichlet type.''  In my opinion, they are of Dirichlet
type only because one started with `Dirichlet type' boundary conditions
(neglecting the questions of needed regularity at the boundary) when one
projected by means of multiplication by $\chi (\mathcal E)$.  
Stated another way:
you can have any kind of selfadjoint boundary conditions on
the surface of the counter, provided you `project' to $\mathcal L^2 (\mathcal
E) \cap \mathcal D (H_0)$ where $\mathcal D (H_0)$ now contains those boundary
conditions.  Then the above evolution-infinitesimal generator commutation
property assures you of a Zeno dynamics in the counter thereafter.

\subsection{Domains and Zeno Measurors}
In my alternate formulation (Section 3), we were able to give conditions 
under which
the `projected' infinitesimal generator $EHE$ remained a generator.  To do so
I needed to let $E$ be a `wider' measuror $A$, so that $AHA$ remained
selfadjoint in the original Hilbert space.  It is desirable, I think, to be able
to remain in the original Hilbert space, in contrast to what I have shown
above to be a reduction of physical description to a unitary evolution trapped
in a counter.

Therefore here I want to return to the approach of [4], viz. Lemma 3.1 above,
and ask, rather than when $(AB)^* = B^* A^*$, the question:  when is $(ABC)^*
= C^* B^* A^*$?  This can then be specialized to:  when is $AHA$ selfadjoint?
By the same analysis as [4] one can establish the following.

\begin{theorem}
Let $A,B,C$ be densely defined operators in a Hilbert space $\mathcal H$.
Suppose the domains $\mathcal D (BC)$ and $\mathcal D (ABC)$ are dense, ranges
$\mathcal R (BC) \supset \mathcal D (A)$, $\mathcal R (C) = \mathcal D (B)$,
domains $\mathcal D ((BC)^*) \supset \mathcal R (A^*)$, $\mathcal D (C^*)
\supset \mathcal R (B^*)$, and $C$ and $BC$ are 1--1.  Then $(ABC)^* = C^* B^*
A^*$.
\end{theorem}

\begin{corollary}
Let $A$ and $H$ be selfadjoint operators in a Hilbert space $\mathcal H$.
Suppose $\mathcal D (HA)$ and $\mathcal D (AHA)$ are dense, $\mathcal R (HA)
\supset \mathcal D (A)$, $\mathcal R (A) \supset \mathcal D (H)$, $\mathcal D
((HA)^*) \supset \mathcal R (A)$, $\mathcal D (A) \supset \mathcal R (H)$.
Then $(AHA)^* = AHA$.
\end{corollary}

The assumptions in Corollary 4.1, although appearing somewhat
general, may be seen to reduce to rather precise statements about domains and
ranges.  I will state this as a further corollary.

\begin{corollary}
Let $A$ and $H$ be selfadjoint operators in a Hilbert space $\mathcal H$.
Then the sufficient conditions of Corollary 4.1 require necessarily that
$\mathcal R (H) = \mathcal D (A)$.
\end{corollary}

\noindent
I remark that the quantum mechanical free space $H_0$ and Poisson--Dirichlet
$H_0$ Hamiltonians we discussed above differ in this respect.  Because 0
is in the continuous spectrum (e.g. see [22]) of the free space $H_0$, we know
that $\mathcal R (H_0)$ is properly dense in $\mathcal L^2 (R^n)$.  So the
measuring observable $A$ should also be unbounded in that case.  The
Poisson--Dirichlet $H_0$ maps onto $\mathcal L^2 (\mathcal E)$ and so the
measuring observable $A$ should be everywhere-defined, and hence bounded, in
that case.

I will give a more complete accounting of these domain considerations
elsewhere.  For example, one may obtain another version of Theorem 4.1 by
factoring $ABC$ in the other way.  There are versions where we may just 
obtain $AHA$ symmetric.  However
I would like again to emphasize the point:  these domain considerations,
essential to selfadjointness and hence for application to obtain 
reversible Zeno
evolutions, are delicate and (in my opinion) should be treated with care in
measurement theory.  In such a way they may provide useful insight into how
measurors should measure.  Thus one may regard my approach as a new
viewpoint towards quantum measurement theory.

\end{document}